\documentclass{epl}

\newcommand{\Obs}{\sigma}

\title{Output spectrum of a measuring device at arbitrary voltage
and temperature}
\shorttitle{Output spectrum ...}
\author{A. Shnirman\inst{1} \and D. Mozyrsky\inst{2}
\and I. Martin\inst{2}}
\institute{
  \inst{1} Institut f\"ur Theor. Festk\"orperphysik,
           Uni. Karlsruhe, D-76128 Karlsruhe, Germany.\\
  \inst{2} Theoretical Division, Los Alamos National Laboratory,
           Los Alamos, NM 87545, USA.
}


\begin{document}

\maketitle

\begin{abstract}
We calculate the noise spectrum of the electrical current in a
quantum point contact which is used for continuous measurements of
a two-level system (qubit). We generalize the previous results
obtained for the regime of high transport voltages (when $V$ is
much larger than the qubit's energy level splitting $B$ (we put
$e=\hbar=1$)) to the case of arbitrary voltages and temperatures.
When $V \sim B$ the background output spectrum is essentially
asymmetric in frequency, i.e., it is no longer classical. Yet, the
spectrum of the amplified signal, i.e., the two coherent peaks at
$\omega=\pm B$ is still symmetric. In the emission (negative
frequency) part of the spectrum the coherent peak can be 8 times
higher than the background pedestal. Alternatively, this ratio 
can be seen in the directly measureable {\it excess} noise. 
For $V < B$ and $T=0$ the coherent peaks do not
appear at all. We relate these results to the properties of linear
amplifiers.
\end{abstract}

\section{Introduction}
\label{sec:Introduction}

For quantum information technology it is necessary to
investigate properties of real physical systems used as quantum
detectors. Certain quantum algorithms require an efficient
(single-shot) read out the final state of a qubit. This
can be done by either strongly coupled threshold
detectors (see e.g., Refs.~\cite{Saclay_Manipulation_Science,Delft_Rabi}),
or by ``measurement in stages'' strategy~\cite{Nakamura_SingleShot}.
For weakly coupled detectors the only way to perform single-shot
measurements is to be in the qunatum-non-demolition (QND) regime,
i.e., by measuring an observable which commutes with the Hamiltonian
and is, thus, conserved. In the solid state domain this regime
has been investigated in, e.g.,
Refs.~\cite{Our_PRB,Clerk_Efficiency,Devoret_Schoelkopf_Nature}.

In this letter we concentrate on continuous weak
non-QND measurements (monitoring) of the coherent oscillations
of a qubit (two-level system, spin-1/2). This regime was the main focus of
Refs.~\cite{Averin_Korotkov,Korotkov_Osc,Averin_SQUID,Ruskov_Korotkov}.
It is realized, e.g.,
for the transverse coupling between the spin and the meter, e.g.,
when the effective magnetic field acting on the spin is along the
$x$-axis while $\sigma_z$ is being measured. In this case one
observes the stationary state properties of the system, after the
information about the initial state of the qubit is lost. Thus,
this regime is not useful for quantum computation.
Yet, studying the properties of the meter in the stationary monitoring
regime, one can obtain information nessecary in order to, later, employ
the meter in the QND regime. Another motivation for our study comes from
the recent activity in the STM single spin detection
(see, e.g., Ref.~\cite{Balatsky_Martin_STM,Bulaevskii_Ortiz}).

Without monitoring and
without coupling to other sources of dissipation the observable
$\sigma_z$ would show coherent (Larmor) oscillations.
When subject to monitoring these oscillations give rise
to a peak in the output spectrum of the meter at the Larmor
frequency.
The laws of quantum mechanics limit the possible height of the peak.
In the case of a 100\% efficient (quantum limited) detector and
when all the noises are white on the frequency scale $B$ the
peak can be only 4 times higher than the background noise pedestal
\cite{Averin_Korotkov}. Inefficiency of the detector reduces the
height of the peak further.

Usually the analysis of the continuous measurements in voltage
driven meters is limited to the case $V\gg
B$~\cite{Averin_Korotkov,Korotkov_Osc,Our_PRB}. The output noise
spectrum in this regime is almost symmetric (classical) at
frequencies of order and smaller than $B$. In this letter we
remove the restriction $V \gg B$. At low voltages, $V \sim B$, the
output noise is essentially asymmetric, i.e., the output signal is
quantum. In other words, we have to differentiate between the
absorption ($\omega >0$) and the emission ($\omega<0$) spectra of
the detector (see, e.g. Ref.~\cite{Gardiner_book}). We show,
however, that the qubit's contribution to the full output noise as
well as to the experimentally accessible {\it excess} noise is
symmetric. We calculate this contribution for arbitrary voltage
and temperature. In the excess noise, which is obtained by
subtracting the equilibrium detector noise ($V=0$) from the output
at $V \ne 0$, the peak to background ratio can reach 8 for $V \sim
B$.

\section{General considerations}
\label{sec:General}

We start from the general theory of linear amplifiers
\cite{Braginsky,Averin_SQUID,Devoret_Schoelkopf_Nature,%
Clerk_Efficiency} which applies in the regime of weak continuous
monitoring. The Hamiltonian of the whole system including the
amplifier (meter) reads $H=H_{\rm meter} + H_{\rm qs} + c \Obs Q$,
where $\Obs$ is the measured observable of the small quantum
system governed by $H_{\rm qs}$, $Q$ is the input variable of the
amplifier governed by $H_{\rm meter}$, and $c$ is the coupling
constant. The meter is necessarily driven out of equilibrium. We
study the output variable of the meter $I$. The stationary average
value $\langle I \rangle$ is only slightly changed by the presence
of the qubit. Much more interesting is the spectrum of
fluctuations $\langle \delta I^2_{\omega}\rangle \equiv \int dt\,
\langle \delta I(t) \delta I(0) \rangle\, e^{i\omega t}$. While it
is convenient to discuss physics in terms of the symmetrized
$S_I(\omega) \equiv (1/2)[\langle \delta I^2_{\omega}\rangle +
\langle \delta I^2_{-\omega}\rangle]$ and anti-symmetrized
$A_I(\omega) \equiv (1/2)[\langle \delta I^2_{\omega}\rangle -
\langle \delta I^2_{-\omega}\rangle]$ correlators, the
calculations are more convenient in terms of the
Keldysh-time-ordered Green's functions (see, e.g.,
Ref.~\cite{Rammer_Smith}). Thus we define $G_{I}(t,t') = -i\langle
T_{\rm K} \delta I(t) \delta I(t')\rangle$. This is a $2 \times 2$
matrix as both $t$ and $t'$ can belong either to the forward or to
the backward Keldysh contours~\cite{Rammer_Smith}. We have the two
basic components $i G_I^{>}(t-t') = \langle \delta I(t) \delta
I(t') \rangle$ and $i G_I^{<}(t-t') = \langle \delta I(t') \delta
I(t) \rangle$ from which all others are built. The retarded and
advanced components are defined as $G_I^{\rm R}(t-t') =
\theta(t'-t) [G_I^{>}(t-t') - G_I^{<}(t-t')]$ and $G_I^{\rm
A}(t-t') = -\theta(t-t') [G_I^{>}(t-t') - G_I^{<}(t-t')]$. These
two components describe, usually, the response of $I$ to a
perturbation coupled to $I$. The Keldysh component defined as
$G_I^{\rm K}(t-t') = G_I^{>}(t-t') + G_I^{<}(t-t')$ is related to
the symmetrized correlator. It is easy to obtain the following
relations: $i G_I^{\rm R}(\omega) - i G_I^{\rm A}(\omega) = 2
A_I(\omega)$ and $i G_I^{\rm K}(\omega) = 2 S_I(\omega)$. We will
also need $G_{IQ}(t,t') \equiv -i \langle T_{\rm K} \delta I(t)
\delta Q(t')\rangle$, $G_{QI}(t,t') \equiv -i \langle T_{\rm K}
\delta Q(t) \delta I(t')\rangle$, and $\Pi(t,t')\equiv -i\langle
T_{\rm K} \delta \Obs(\tau) \delta \Obs(\tau')\rangle$. Various
components of these Green's functions are defined analogously to
those of $G_I$.

We assume that one is allowed to use
Wick's theorem for the operators $I$, $Q$, and $\Obs$.
Frequently, even if Wick's theorem does not apply, one can still
use it for the lowest (in the coupling constant) calculations
and show that the corrections are of the higher order.
We return to this subject later.
Since the dynamics of the
measured system changes substantially as a result of measurement
while the meter's one is perturbed weakly, we use the full
(``thick'') Green's function $\Pi$, while for $G_{IQ}$ and
$G_{QI}$ one keeps the unperturbed values.
Then the lowest order irreducible correction to the Green's
function $G_{I}(t,t')$ reads
\begin{equation}
\label{Eq:dG_I} \delta G_{I}(t,t') = c^2\oint\oint d\tau d\tau'
\,G_{IQ}(t,\tau)\,\Pi(\tau,\tau')\,G_{QI}(\tau',t') \ .
\end{equation}
In the stationary regime this gives
\begin{equation}
\label{Eq:dG_I_Matrix} \delta G_I(\omega) = c^2\left(
\begin{array}{cc}
G_{IQ}^{\rm R}(\omega) & G_{IQ}^{\rm K}(\omega)\\
0 & G_{IQ}^{\rm A}(\omega)
\end{array}
\right) \left(
\begin{array}{cc}
\Pi^{\rm R}(\omega) & \Pi^{\rm K}(\omega)\\
0 & \Pi^{\rm A}(\omega)
\end{array}
\right) \left(
\begin{array}{cc}
G_{QI}^{\rm R}(\omega) & G_{QI}^{\rm K}(\omega)\\
0 & G_{QI}^{\rm A}(\omega)
\end{array}
\right) \ .
\end{equation}
For the Green's function $G_{QI}$ we have
$G_{QI}^{\rm R}(\omega) = G_{IQ}^{\rm A}(-\omega)$, $G_{QI}^{\rm A}(\omega) =
G_{IQ}^{\rm R}(-\omega)$, and $G_{QI}^{\rm K}(\omega) =
G_{IQ}^{\rm K}(-\omega)$. We introduce the notations $\lambda(\omega)
= c\,G_{IQ}^{\rm R}(\omega)$, $\lambda'(\omega) =
c\,G_{QI}^{\rm R}(\omega)$, where $\lambda$ is the direct gain
(amplification coefficient) of the amplifier, while $\lambda'$ is
the inverse gain. As $\lambda(t)$ and $\lambda'(t)$ are real,
$\lambda(-\omega) = \lambda^{*}(\omega)$ and $\lambda'(-\omega) =
\lambda'^{*}(\omega)$. Thus we obtain
\begin{eqnarray}
\label{Eq:dGR}
\delta G_{I}^{\rm R}(\omega) &=& \lambda(\omega) \lambda'(\omega)
\Pi^{\rm R}(\omega)\ ,\\ \label{Eq:dGK}\delta G_{I}^{\rm
K}(\omega) = -2i \delta S_I(\omega) &=& |\lambda(\omega)|^2\,
\Pi^{\rm K}(\omega)+2ic\,
{\rm Im}\left[\lambda(\omega)\,\Pi^{\rm
R}(\omega)\,G_{QI}^{\rm K}(\omega)\right] \ ,
\end{eqnarray}
and $\delta G_{I}^{\rm A}(\omega) = [\delta G_{I}^{\rm
R}(\omega)]^{*}$. We have also used $G_{QI}^{\rm K}(\omega) =
G_{IQ}^{\rm K}(-\omega)=-[G_{IQ}^{\rm K}(\omega)]^*$.
The first term of
(\ref{Eq:dGK}) corresponds to the noise of
the small system "amplified" by the meter. The second term is needed
to fulfill the fluctuations-dissipation relation at equilibrium.
We will see that it is also important at low voltages, i.e.,
when the detector is not driven far enough from equilibrium.
For good amplifiers the inverse gain vanishes, $\lambda'=0$, and
we obtain $\delta G_I^{\rm R}=0$. Thus the contribution to the output correlator
$\delta\, \langle \delta I^2_{\omega}\rangle = i\delta
G_{I}^{>}=(i/2)(\delta G_{I}^{\rm K}+\delta G_{I}^{\rm R}-\delta
G_{I}^{\rm A})= (i/2)\delta G_{I}^{\rm K} = \delta S_I(\omega)$ is
symmetric in frequency, i.e., $\delta A_I(\omega) = 0$.
Vanishing of $\lambda'$ also means that further
amplifiers using $I$ as an input will not add to the back-action.

\section{Spin's dynamics}
\label{sec:Spin}

When an observable of a qubit (a component of spin-1/2) is being measured
we can assume without loss of generality $\Obs=\sigma_z$. The spin is
subject to an (effective) magnetic field $\vec{B}$, i.e.,
$H_{\rm qs} = -(1/2)\vec{B}\,\vec{\sigma}$. Its dynamics
is, thus, obtained from the Hamiltonian
\begin{equation}
H= -\frac{1}{2}\,\vec{B}\,\vec{\sigma} - \frac{1}{2}\,Q \sigma_z
\ ,
\end{equation}
where we have put $c=-1/2$ so that $Q$ can be interpreted as
fluctuating magnetic field. We exclude the case
$\vec{B} \parallel \bf{z}$, in which the measured observable
$\Obs=\sigma_z$ commutes with the Hamiltonian and, thus, is conserved.
This regime is known as the quantum-non-demolition (QND) one and
has been treated, e.g., in
Refs.~\cite{Our_PRB,Clerk_Efficiency,Devoret_Schoelkopf_Nature}.
In all other cases the stationary state is reached after some
transient period and we can study the
output spectrum of the meter.

For simplicity we assume no extra dissipation sources acting on the qubit
except for the meter. The spin's Green functions
(correlators) are obtained within the standard Bloch-Redfield
approach~\cite{Bloch_Derivation,Redfield_Derivation} which is applicable
as long as the dissipation is weak (see below). Within this
approach one, first, calculates the markovian evolution operator
for the spin's density matrix, and, then, uses the ``quantum
regression theorem'' \cite{Gardiner_book} to obtain the correlators.
For this lowest order perturbative (in the spin-meter coupling)
calculation one only needs to know the (unperturbed by the spin)
fluctuations spectrum, $\langle Q^2_{\omega}\rangle$.
One, then, obtains
\begin{eqnarray}
\label{Eq:P_KRA}
&&\Pi^{\rm K}(\omega) =
-i\sin^2\theta \left[\frac{2\Gamma_2}{(\omega-B)^2+\Gamma_2^2}
+\frac{2\Gamma_2}{(\omega+B)^2+\Gamma_2^2}
\right]
-i\cos^2\theta\,\frac{4\Gamma_1}{\omega^2+\Gamma_1^2}\,
\left[1-\langle \sigma_{\vec{B}} \rangle^2\right]
\ ,\nonumber \\
&&\Pi^{\rm R}(\omega) =
\sin^2\theta\,\langle \sigma_{\vec{B}} \rangle\,
\left[\frac{1}{\omega-B+i\Gamma_2}
-\frac{1}{\omega+B+i\Gamma_2}
\right]
\ ,
\end{eqnarray}
where $\theta$ is the angle between $\vec{B}$ and $\bf{z}$.
The stationary spin polarization along $\vec{B}$ is given by
$\langle \sigma_{\vec{B}} \rangle = h(B)$, where $h(\omega)\equiv
A_Q(\omega)/S_Q(\omega)$, while
$S_Q(\omega) \equiv (1/2)[\langle Q^2_{\omega}\rangle +
\langle Q^2_{-\omega}\rangle]$ and $A_Q(\omega) \equiv
(1/2)[\langle Q^2_{\omega}\rangle - \langle
Q^2_{-\omega}\rangle]$.
The relaxation ($\Gamma_1$) and the dephasing ($\Gamma_2$)
rates are given by:
\begin{equation}
\Gamma_1 = (1/2)\sin^2\theta\,S_Q(\omega=B) \ \ \ , \ \ \
\Gamma_2=(1/2)\Gamma_1 + (1/2)\cos^2\theta\,S_Q(\omega=0)
\ .
\end{equation}
The applicability condition of the Bloch-Redfield approach is
$\Gamma_1,\Gamma_2,\delta B \ll B$, where $\delta B$ is the
renormalization of the energy splitting (Lamb shift). If one
treats the Lorentzians in (\ref{Eq:P_KRA}) as true delta
functions, one can derive from the second equation of
(\ref{Eq:P_KRA}) a relation resembling the fluctuation-dissipation
theorem: $\Pi^{\rm R}(\omega)-\Pi^{\rm A}(\omega) \approx
h(\omega)\,\Pi^{\rm K}(\omega)$. 
Note, that, although we use the ``diagrammatic'' language
of Keldysh Green functions, Eqs.~(\ref{Eq:P_KRA}) are obtained
without any diagrams or assumptions about the applicability of
Wick's theorem.

\section{Spin's contribution to the output spectrum}

Substituting Eqs.~(\ref{Eq:P_KRA}) into Eq.~(\ref{Eq:dGK})
we obtain spin's contribution to the symmetrized output spectrum
of the meter:
\begin{eqnarray}
\label{Eq:delta_S_I_general}
\delta S_I(\omega>0) &=& \sin^2\theta \left[\frac{\Gamma_2}{(\omega-B)^2+\Gamma_2^2}
\right] \left\{|\lambda(\omega)|^2 -\frac{h(B)}{2}\, {\rm Re}\left[\lambda(\omega)
G_{QI}^{\rm K}(\omega)\right]\right\}
\nonumber\\
&+&  \sin^2\theta \left[\frac{\omega-B}{(\omega-B)^2+\Gamma_2^2}
\right]\, \frac{h(B)}{2}\, {\rm Im}\left[\lambda(\omega)
G_{QI}^{\rm K}(\omega)\right]
\nonumber\\
&+& \cos^2\theta\,\frac{2\Gamma_1}{\omega^2+\Gamma_1^2}\,
\left[1-h^2(B)\right]\,|\lambda(\omega)|^2
\ .
\end{eqnarray}
We assume that $\lambda(\omega)$ and $G_{QI}^{\rm K}(\omega)$ are smooth
near $\omega = \pm B$ and
$\omega =0$. Then, the first term of (\ref{Eq:delta_S_I_general})
gives two peaks near $\omega = \pm B$ with width $\Gamma_2$.
The third term gives a peak at
$\omega = 0$ with widt $\Gamma_1$.
The second term of (\ref{Eq:delta_S_I_general})
gives the Fano shaped contributions near $\omega = \pm B$.
The simplest situation arises when $\theta = \pi/2$
and $h(B) \rightarrow 0$ (at very high transport
voltages the effective temperature of the spin is infinite and
$\langle \sigma_{\vec B}\rangle \rightarrow 0$). Then only the peaks
at $\omega = \pm B$ survive with the height $\delta S_I(B) = |\lambda(B)|^2/\Gamma_2=
4|\lambda(B)|^2/S_Q(B)$. The peak to pedestal ratio
$\delta S_I(B)/S_I(B) = 4|\lambda(B)|^2/(S_Q(B)S_I(B))$ was
shown~\cite{Averin_Korotkov,Korotkov_Osc,Averin_SQUID,Ruskov_Korotkov}
to be limited by 4. Below we investigate the coherent peaks at arbitrary voltage and
temperature for a specific example of a meter with the purpose to explore the effect
of the rest of the terms in (\ref{Eq:delta_S_I_general}).

\section{Quantum Point Contact (QPC) as a meter}
\label{sec:System}

The QPC devices are known to serve as effective
meters of charge (see, e.g.,
Refs.~\cite{Field,Sprinzak_Charge,Buks,%
Kouwenhoven_Charge}).
The conductance of the QPC is controlled by the quantum state of a
qubit. We focus on the simplest limit of a tunnel junction when
the transmissions of all the transport channels are much smaller
than unity. This model has previously been used by many
authors~\cite{Gurvitz,Korotkov_Continuous,Goan_Dynamics}.
It also corresponds to the model considered
in Ref.~\cite{Bulaevskii_Ortiz} in the regime of lead electrons
fully polarized along the $z$ axis,
$\bf{m}_{\rm R} = \bf{m}_{\rm L} = \bf{z}$.
The tunnel junction limit is described by the following
Hamiltonian
\begin{eqnarray}
\label{eq:Hamiltonian} H = \sum_{l} \epsilon_{l}
    c_{l}^{\dag}\,c_{l}^{\phantom\dag} +
    \sum_{r} \epsilon_{r} c_{r}^{\dag}\,c_{r}^{\phantom\dag} +
    \sum_{l,r} (t_0+ t_1\sigma_z) (c_{r}^{\dag}\,c_{l}^{\phantom\dag} + h.c.)
     -(1/2)\,\vec{B}\,\vec{\sigma}
\ .
\end{eqnarray}
The transmission
amplitudes $t_0$
and $t_1$ are assumed to be real positive and small
(tunnel junction limit). We also assume $t_1 \ll t_0$ to be in the
linear amplifier regime.

For brevity we introduce the operator $X\equiv
\sum_{l,r}c_{l}^{\dag}\,c_{r}^{\phantom\dag}$ and then $j\equiv
i(X-X^{\dag})$ and $q\equiv (X+X^{\dag})$. The current operator is
given by $I=(t_0+t_1\sigma_z) j$, while the tunneling Hamiltonian
is $H_{\rm T} = (t_0+t_1\sigma_z) q$. We
see that the analysis of the amplifiers presented above cannot be
directly applied. First, the interaction term between the spin and
the QPC, i.e., $t_1\, q\, \sigma_z$ (thus, in our case $Q=-2 t_1
q$), is not the full interaction vertex but rather a part of
$H_{\rm T}$. Second, the current operator $I$ contains the spin's
operator $\sigma_z$ explicitly. One possible way to resolve these
difficulties (see e.g., \cite{Averin_Korotkov,Averin_SQUID}) is to
include the spin-independent part of $H_{\rm T}$, namely $t_0\,
q$, into the zeroth-order Hamiltonian. This amounts to working in the
basis of scattering states. Here we adopt a simpler procedure suitable for
QPC's in the tunneling regime. We expand in the full $H_{\rm T}$
and keep all the terms up to the order $t_0^2\,t_1^2$. For this we
need the following zeroth-order Green's functions: $G_{qq} \equiv
-i\langle T_{\rm K}q(t)q(t')\rangle$, $G_{jj} \equiv -i\langle
T_{\rm K}j(t)j(t')\rangle$, and $G_{jq} \equiv -i\langle
T_{\rm K}j(t)q(t')\rangle$, $G_{qj} \equiv -i\langle
T_{\rm K}q(t)j(t')\rangle$.
For $\omega \ll D$, where $D$ is the
electronic bandwidth (the Fermi energy) we obtain
\begin{equation}
\label{Eq:G_qq}
G_{qq}(\omega)=G_{jj}(\omega)=-i\eta
\left(\begin{array}{cc}
\omega + i... & 2 s(\omega) \\
0 & -\omega + i...
\end{array}\right)
\ ,
\end{equation}
\begin{equation}
\label{Eq:G_jq}
G_{jq}(\omega)=-G_{qj}(\omega)=\eta \left(\begin{array}{cc}
V(1 + iO(\omega/D)) & 2 a(\omega) \\
0 & -V(1 - iO(\omega/D))
\end{array}\right)
\ ,
\end{equation}
where $\eta \equiv 2\pi\rho_{\rm L}\rho_{\rm R}$. We have also
introduced the two following functions:
\begin{eqnarray}
s/a\,(\omega)\equiv
\frac{V+\omega}{2}\,\coth\frac{V+\omega}{2T}\pm
\frac{V-\omega}{2}\,\coth\frac{V-\omega}{2T} \ .
\end{eqnarray}
In Eq.~(\ref{Eq:G_qq}) $...$ stand for the real part of the
retarded (advanced) components. The factors $1\pm iO(\omega/D)$ in
(\ref{Eq:G_jq}) are responsible for making the functions
$G_{jq}^{\rm R}(t)$ and $G_{jq}^{\rm A}(t)$ causal. As we are
interested in the low frequencies ($\omega \ll D$) we approximate
those factors by $1$.

\section{Peaks in the output noise spectrum}
\label{Sec:Peaks}

\begin{figure}
\onefigure[width=10cm]{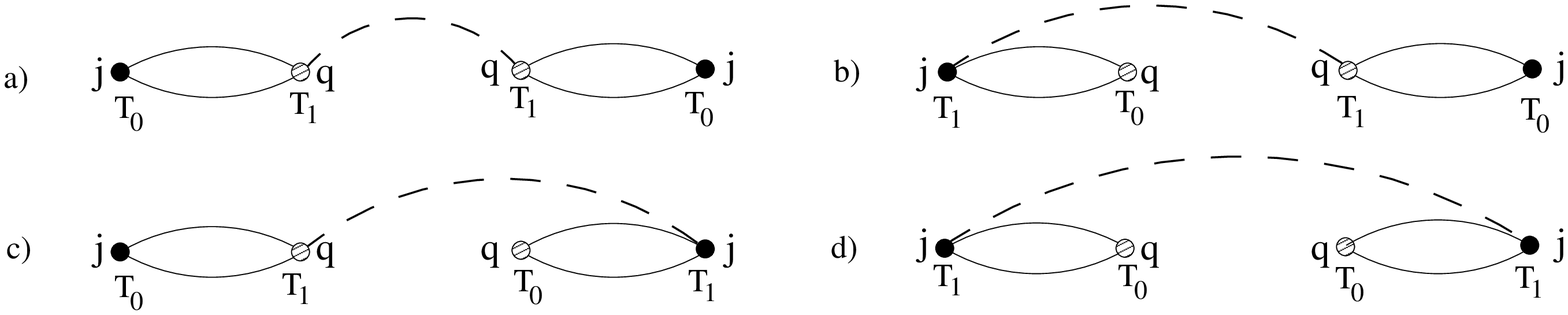} \caption{The diagrams
leading to Eq.~(\ref{Eq:dG_I_QPC}).
The loops represent the Green functions $G_{jq}$ and $G_{qj}$.
Each of these Green's
functions is actually a combination of two electronic ones
(solid lines). The dashed line is the spin's Green function.}
\label{Figure:Keldysh_Peak}
\end{figure}
\begin{figure}
\twofigures[width=4.5cm]{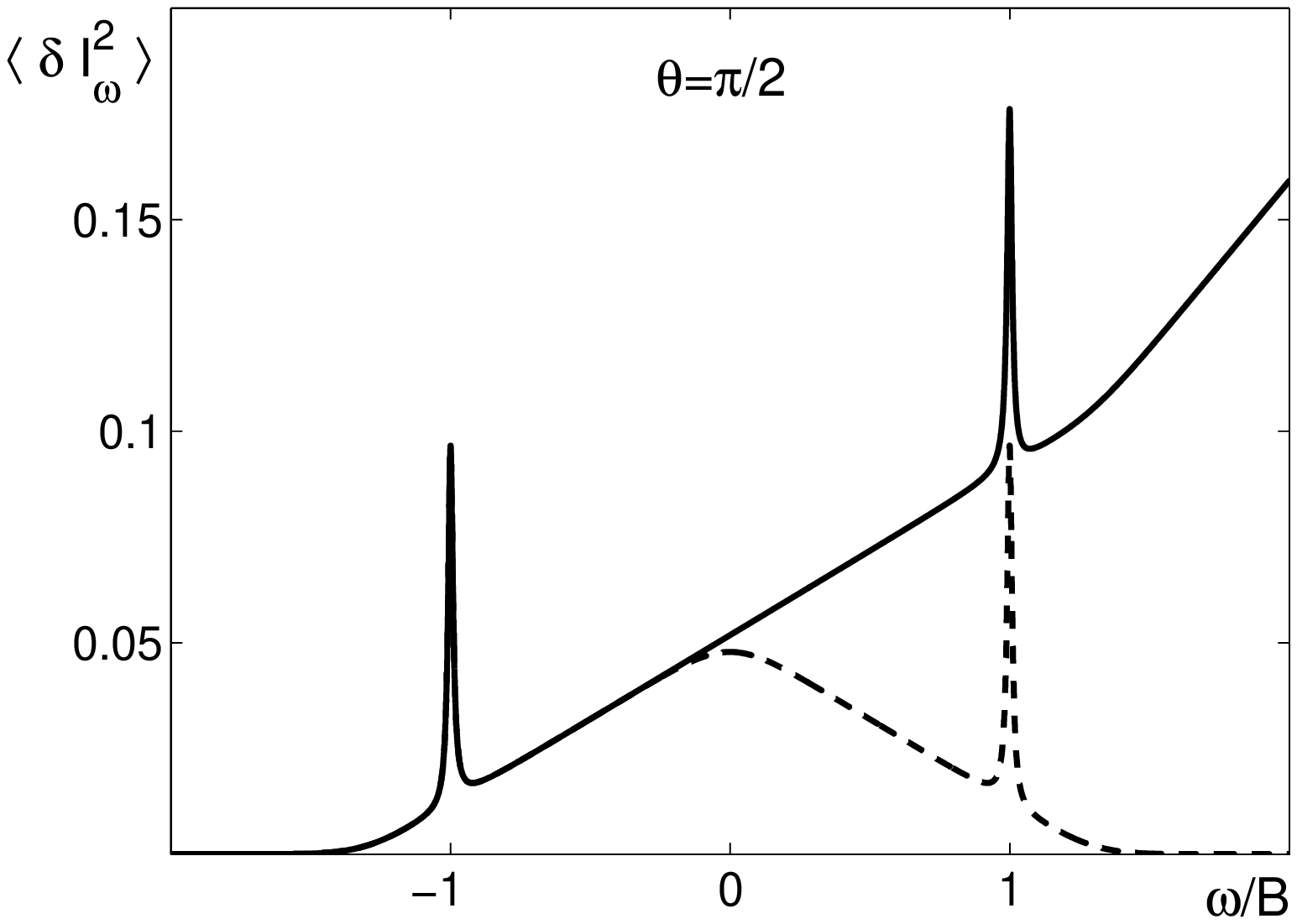}{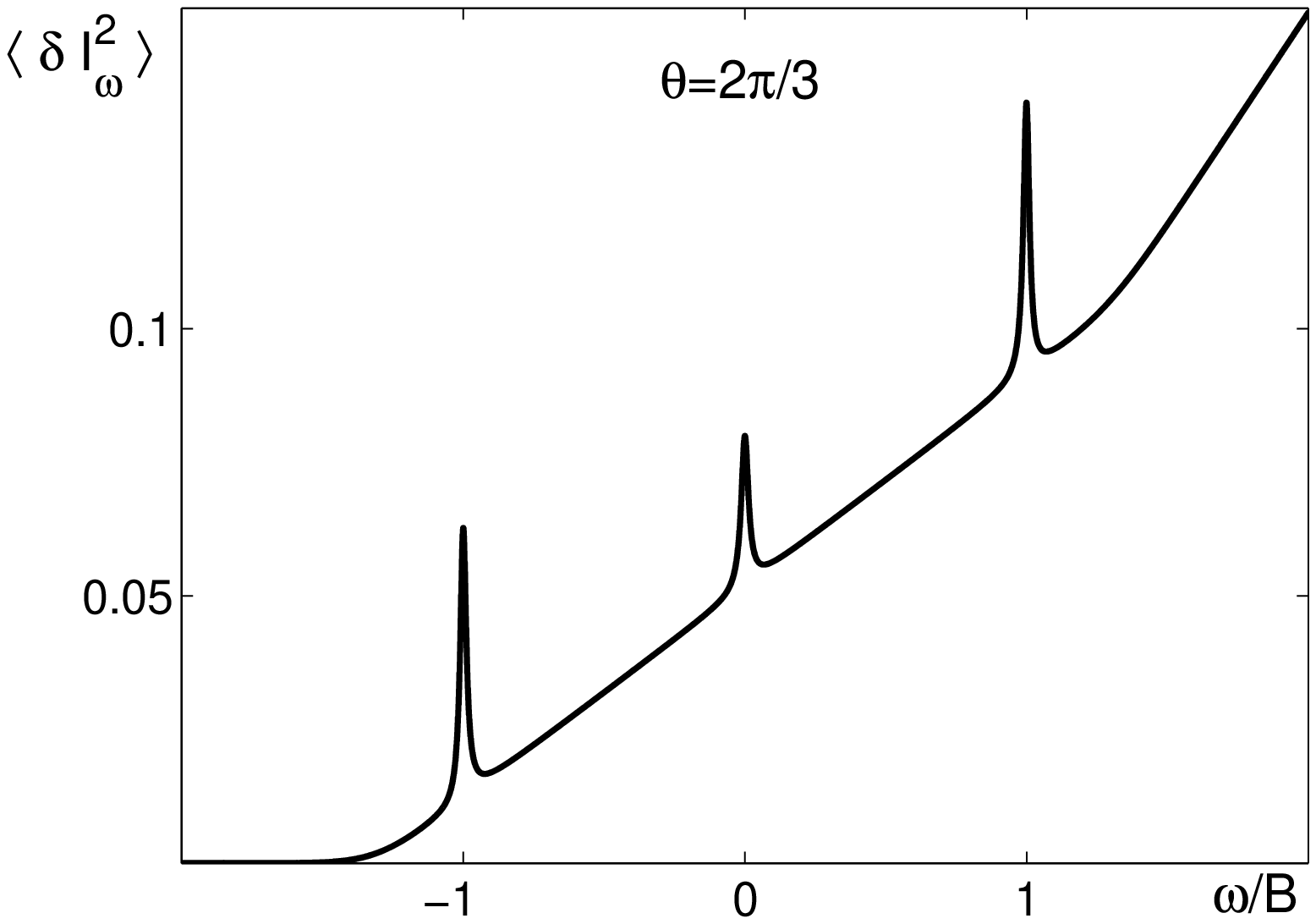}
\caption{Output noise power for $\theta=\pi/2$, $g_0=0.25$,
$g_1=0.05$, $T=0.05B$, and $V=1.3B$. Solid line: the
full correlator; dashed line: the excess spectrum.}
\label{Figure:pi2}
\caption{Same parameters but $\theta=2\pi/3$. Only
the full correlator shown.}
\label{Figure:2pi3}
\end{figure}

We combine the Green functions
$G_{jq}(\omega)$, $G_{qj}(\omega)$, and $\Pi(\omega)$
(see Eq.~\ref{Eq:P_KRA}) into diagrams to calculate
the qubit's contribution to the current-current Green
function $\delta G_I$.
The Wick theorem does not apply to the spin operators. However,
using the Majorana representation of the spin operators and recently proved
useful identities (see Refs.~\cite{Coleman_Identity,Shnirman_Makhlin_Identity})
we are able to show that in order $t_0^2 t_1^2$ the answer is given by
the diagrams shown in Fig.~\ref{Figure:Keldysh_Peak}. We obtain
$\delta G_I =t_0^2 t_1^2
\left[G_{jq}(\omega)+G_{jq}^{\rm R}(0) \cdot \hat 1\right]\;
\Pi(\omega)\;\left[G_{qj}(\omega) + G_{qj}^{\rm A}(0)\cdot \hat 1\right]
$,
which can be rewritten as
\begin{equation}
\label{Eq:dG_I_QPC} \delta G_I = (1/\pi^{2})\, g_0 g_1 \left(
\begin{array}{cc}
V & a(\omega)\\
0 & 0
\end{array}
\right) \left(
\begin{array}{cc}
\Pi^{\rm R} & \Pi^{\rm K}\\
0 & \Pi^{\rm A}
\end{array}
\right) \left(
\begin{array}{cc}
0 & -a(\omega)\\
0 & V
\end{array}
\right) \ ,
\end{equation}
where
we have defined the conductances as $g_0\equiv 2\pi \eta t_0^2$ and
$g_1\equiv 2\pi \eta t_1^2$
It is worth comparing Eqs.~(\ref{Eq:dG_I_QPC}) and (\ref{Eq:dG_I_Matrix}).
Even though the simple formalism leading to
Eq.~(\ref{Eq:dG_I_Matrix}) was not directly
applicable in our case, the result (\ref{Eq:dG_I_QPC})
looks very similar. We can
interpret, therefore, $\lambda = (1/\pi)\,\sqrt{g_0 g_1}V$, $\lambda'=0$, and
$c G_{IQ}^{\rm K}(\omega) = (1/\pi)\,\sqrt{g_0 g_1}\,a(\omega)$.
Thus the tunnel barrier possesses the property $\lambda'=0$ at all
frequencies. This assures that $\delta G^{\rm R/A}_I=0$ and the
contribution to the current-current correlator is symmetric:
\begin{eqnarray}
\label{Eq:d_S_QPC} \delta S_I(\omega>0) &=&
(1/\pi^2)\, g_0 g_1 V^2
\sin^2\theta\;\frac{\Gamma_2}{(\omega-B)^2+\Gamma_2^2}\;
\left[1-\frac{a(\omega)h(B)}{V}\right]
\nonumber \\
&+& (1/\pi^2)\, g_0 g_1 V^2 \cos^2\theta\;\frac{2\Gamma_1}{\omega^2+\Gamma_1^2}\;
\left[1-h^2(B)\right]
\ .
\end{eqnarray}
We note that in our example $S_Q(\omega) =
2 i t_1^2 G_{qq}^{\rm K}(\omega) = (2/\pi)\, g_1 s(\omega)$ and
$A_Q(\omega) = (2/\pi)\, g_1\omega$. Then we obtain
$h(B) = B/s(B)$, $\Gamma_1 = (1/\pi)\, g_1 \sin^2\theta\; s(B)$, and
$\Gamma_2 = (1/2\pi)\,g_1 \sin^2\theta\; s(B) + (1/\pi)\, g_1 \cos^2\theta\; s(0)$.
Note, that no Fano shaped contributions appear due to the
fact that both $\lambda(\omega)$ and $ G_{IQ}^{\rm K}(\omega)$ are
real. The Lorentzians in
Eq.~(\ref{Eq:d_S_QPC}) coincide with the ones
obtained in Refs.~\cite{Averin_Korotkov,Korotkov_Osc,Averin_SQUID,Ruskov_Korotkov}.
The new result is the reduction factor for the peaks at $\omega = \pm B$
in the square brackets. This factor simplifies
for $T=0$. Then, if $V>B$, it is given by $(1-B^2/V^2)$,
while for $V<B$ it is equal to $0$. In the last case the
measuring device can not provide enough energy to excite the qubit
and, therefore, the qubit remains in the ground state and does not
produce any additional noise (see also Ref.~\cite{Bulaevskii_Ortiz}).
The ratio between the peak's height
and the pedestal's height is different for positive and negative
frequencies. In the limit $g_1\ll g_0$ we obtain
$\langle \delta I^2_{\omega}\rangle  \approx it_0^2 G_{jj}^{>}(\omega) =
(1/2\pi)\,g_0(s(\omega)+\omega)$. Thus, for $T=0$, and $B < V$ we
obtain $\langle \delta I^2_{\pm B}\rangle \approx (1/2\pi)\,g_0 V (1 \pm B/V)$
and $\delta \langle \delta I^2_{\pm B}\rangle = \delta S_I(\pm B)
\approx (2/\pi)\,g_0 V (1-B^2/V^2)$
and
\begin{equation}
\frac{\delta \langle \delta I^2_{\pm B}\rangle}{\langle
\delta I^2_{\pm B}\rangle} \approx
4(1\mp\frac{B}{V}) \ .
\end{equation}
For $B \rightarrow V$ the ratio for the negative frequency
peak reaches $8$. In this limit, however, the peak's hight is
zero. For symmetrized spectra the maximal possible ratio is $4$
(Ref.~\cite{Korotkov_Osc}). An interesting question is what
exactly can be observed experimentally. If the further detection
of the output noise is passive, like the photon counting in
fluorescence experiments, one can only measure what the system
emits, i.e. the noise at negative
frequencies~\cite{Lesovik_Loosen,Gavish_Levinson_Imry,Aguado}.
Moreover,
in our example, the {\it excess} noise, i.e., $\langle \delta I^2_{\pm B}\rangle (V) -
\langle \delta I^2_{\pm B}\rangle (V=0)$, is symmetric.
As shown in Ref.~\cite{Gavish_Imry_Levinson_Yurke},
if the excess noise is symmetric, it can be effectively measured
even by a finite temperature LCR filter.
In Figs.~\ref{Figure:pi2},\ref{Figure:2pi3}
we show examples of output noise spectrum and of the
corresponding excess noise spectrum.

\section{Conclusions}
\label{sec:Conclusions}

We have calculated the output noise of the point contact used as a
quantum detector of qubit's coherent oscillations
for arbitrary voltage and temperature. In the
regime $eV\sim B$ and $T\ll B$ the output noise is
essentially asymmetric. Yet, the qubit's oscillations produce two
symmetric peaks at $\omega =\pm B$ and also a peak at $\omega=0$.
Due to the vanishing of the inverse gain ($\lambda'=0$)
the peaks  at $\omega =\pm B$ have equal height and, therefore,
the negative frequency peak is much higher
relative to it's pedestal than the positive frequency one.
The peak/pedestal ratio can reach 8. This
can be observed by further passive detectors, which measure what the
system emits, or by measuring the {\it excess} noise.
The results of this paper are obtained for the
simplest and somewhat artificial model of a quantum detector, a QPC in the
tunneling regime. It would be interesting to perform
analogous calculations for more realistic detectors like SET's or
QPC's with open channels (see e.g.,
Refs.~\cite{Pilgram_Buettiker,Clerk_Efficiency,Aguado}).

\acknowledgments

We thank Yu.~Makhlin, Y. Levinson, and L. Bulaevskii
for numerous fruitful discussions.
A.S. was supported by the EU IST Project SQUBIT and
by the CFN (DFG). D.M. and I.M. were supported by the U.S. DOE.

\bibliographystyle{mybst}
\bibliography{ref}

\end{document}